\documentclass[letterpaper,10pt,twocolumn,final,conference,oneside]{IEEEtran}
\usepackage{array}
\usepackage{epsfig}
\usepackage{graphics}
\usepackage{graphicx}
\usepackage[english]{babel}
\usepackage[latin1]{inputenc}
\usepackage{amsmath}
\usepackage{amssymb}
\usepackage{booktabs}
\usepackage{floatrow}
\usepackage{subfigure}
\usepackage{float}
\usepackage{bm}
\usepackage{cite}
\usepackage{cases}
\usepackage{color,soul}
\usepackage{comment}
\usepackage{multicol}
\usepackage{url}
\usepackage{multirow}
\usepackage{balance}
\usepackage{acronym}
\usepackage{chemformula}
\usepackage{eurosym}
\usepackage{tikz}

\usepackage{siunitx} 
\DeclareSIUnit{\atm}{atm}
\DeclareSIUnit{\kWh}{kWh}
\DeclareSIUnit{\Ah}{Ah}
\DeclareSIUnit{\eur}{\mbox{\text{\euro}}}
\DeclareSIUnit{\kW}{kW}
\DeclareSIUnit{\MW}{MW}
\DeclareSIUnit{\kwm}{kW/m^2}
\DeclareSIUnit{\m}{m}

\restylefloat{table}
\acrodef{bess}[BESS]{Battery Energy Storage System}
\acrodef{ptg}[PtG]{Power to Gas}
\acrodef{ess}[ESS]{Energy Storage System}
\acrodef{sps}[SPS]{shipboard Power System}
\acrodef{fc}[FC]{Fuel Cell}
\acrodef{aes}[AES]{All-Electric Ship}
\acrodef{dg}[DG]{Diesel Generator}
\acrodef{ci}[CI]{Cold Ironing}
\acrodef{res}[RES]{Renewable Energy Sources}
\acrodef{pv}[PV]{Photovoltaic}
\acrodef{mpc}[MPC]{Model Predictive Control}
\acrodef{soc}[SoC]{State of Charge}
\acrodef{soh}[SoH]{Hydrogen Level in the Storage}
\acrodef{sos2}[SOS2]{Special Ordered Set of type 2}
\acrodef{hss}[HESS]{Hydrogen Energy Storage System}
\acrodef{minlp}[MINLP]{Mixed Integer Non-Linear Problem}
\acrodef{miqcp}[MIQCP]{Mixed Integer Quadratic-Constrained Problem}
\acrodef{ze}[ZE-Ships]{Zero-Emission Ships}
\acrodef{fc}[FC]{Fuel Cell}
\acrodef{wt}[WT]{Wind Turbine}
\acrodef{lcoe}[LCOE]{Levelized Cost of Energy}
\acrodef{res}[RES]{Renewable Energy Sources}
\acrodef{el}[Ely]{Electrolyzer}
\acrodef{mes}[MES]{Multi-Energy System}
\acrodef{gams}[GAMS]{General Algebraic Modeling System}
\acrodef{wf}[WF]{Wind Farm}

\IEEEoverridecommandlockouts
\newcommand\copyrighttext{%
  \footnotesize
  \centering\copyright~2022 IEEE. Personal use of this material is permitted. Permission from IEEE must be obtained for all other uses, in any current or future media, including reprinting/republishing this material for advertising or promotional purposes, creating new collective works, for resale or redistribution to servers or lists, or reuse of any copyrighted component of this work in other works. \\ Presented at the IEEE PES General Meeting 2022. DOI: 10.1109/PESGM48719.2022.9916817}
\newcommand\copyrightnotice{%
\begin{tikzpicture}[remember picture,overlay]
\node[anchor=south,yshift=0pt] at (current page.south) {\setlength{\fboxrule}{0pt}\fbox{\parbox{\dimexpr\textwidth-\fboxsep-\fboxrule\relax}{\copyrighttext}}};
\end{tikzpicture}%
}

\begin{document}

\title{Optimal Management of a Smart Port with Shore-Connection and Hydrogen Supplying by Stochastic Model Predictive Control}

\author{%
  \IEEEauthorblockN{F. Conte}
  \IEEEauthorblockA{Campus Bio-Medico University of Rome\\
    Faculty of Engineering\\
    Via Alvaro del Portillo, 21\\
    I-00128, Roma, Italy\\
    f.conte@unicampus.it}
          \and
     \IEEEauthorblockN{F. D'Agostino, D. Kaza, S. Massucco, G. Natrella, F. Silvestro}
 \IEEEauthorblockA{University of Genoa\\
    DITEN\\
    Via all'Opera Pia 11 A\\
    I-16145, Genova, Italy\\
    federico.silvestro@unige.it}
    }

\IEEEaftertitletext{\copyrightnotice\vspace{1.1\baselineskip}}
\maketitle

\IEEEpubidadjcol
\begin{abstract}
The paper proposes an optimal management strategy for a Smart Port equipped with renewable generation and composed by an electrified quay, operating Cold-Ironing, and a Hydrogen-based quay, supplying Zero-Emission Ships. One Battery Energy Storage System and one Hydrogen Energy Storage System are used to manage renewable energy sources and to supply electric and hydrogen-fueled ships. A model predictive control based algorithm is designed to define the best economic strategy to be followed during operations. The control algorithm takes into account the uncertainties of renewable energy generation using stochastic optimization. The performance of the approach is tested on a potential future Smart Port equipped with wind and photovoltaic generation.   
\end{abstract}

\begin{IEEEkeywords}
Smart Port, Hydrogen, Stochastic Model Predictive Control, Cold-Ironing, Multi-Energy Systems.
\end{IEEEkeywords}


\acresetall\acused{ips}
\section{Introduction}\label{sec:Introduction}
The continuous increase of global warming due to pollution has raised the attention in improving the systems in the marine sector. Focusing to Europe, the maritime sector produced 138~million~tons of \ch{CO2} in 2018, which is more than 3\% of the total \ch{CO2} continental emissions  \cite{EUmaritime}. Seagoing vessels are not the only contributors to air pollution since also ships moored at ports produce a significant portion of emissions \cite{John2014}. Indeed, a ship moored at port requires a certain amount of energy to be self-sustained, which is today provided by on-board power stations, usually composed by a set of \acp{dg}.

An efficient way to reduce emissions at berth is to shut down the \acp{dg} and supply ships with a shore-connection. This operation, called \ac{ci}, is recommended by the European Union in \cite{ColdIroning}, where EU ports are required to provide facilities to enable its implementation by 2025.

Literature provides many assessments of the potentialities of \ac{ci}. In \cite{d2019integration} it is shown that \ch{CO2} emissions can be reduced by \ac{ci} in the range of \SIrange{48}{70}{\percent}. In \cite{zis2014evaluation} it is assessed that \ac{ci} may reduce \ch{CO2} and \ch{NO_{X}} emissions by \SI{25}{\percent} and \SI{92}{\percent}, respectively.

The environmental benefits of \ac{ci} is leading ports to build shore-connection infrastructures. There are examples in US, Canada  and  Europe \cite{Kumar2019}. However, the adoption of \ac{ci} is not very widespread due to the economic convenience of producing energy on board compared to purchasing energy from the electric market. To overcome this problem, many solutions have been proposed. In Europe, the first \ac{ci} has been implemented in the port of Gothenburg (Sweden) \cite{ZIS201982}. Here, \ac{ci} has been encouraged by applying an extra tax to all not \ac{ci} users. In Italy, the decree \cite{milleproroghe} applied since 2020 a tax of \SI{0.0005}{\eur/\kW\hour} for ships moored in ports equipped with on-board power systems with rated power above \SI{35}{\kW}.  

In this context, this paper proposes an optimal management strategy for a smart port operating \ac{ci}. The port is also a \ac{mes}, since it is provided with \ac{res} coupled with a \ac{bess} and performs hydrogen supplying. Indeed, an alternative way to reduce ship emissions, which is receiving particular attention \cite{banaeioptimal}, is to realize \ac{ze}, where \acp{dg} are substituted by \acp{fc}.
Therefore, the idea of this article is to supply \ac{ze} at berth with the hydrogen needed to produce self-sustaining energy. This hydrogen is self-produced in the smart port using the available \ac{res} (green hydrogen) by an \ac{el} that, together with a \ac{fc} and a hydrogen storage, composes a \ac{hss}. 

The management of this hypothetical future smart port, is realized by a stochastic \ac{mpc} algorithm. Literature provides several uses of \ac{mpc} in smart cities with \ac{ptg} devices \cite{Fischer:2018} or, in general, for \ac{mes} \cite{Guo:2019}. The peculiarity of the proposed approach is that uncertainties of \ac{res} are considered to maximize the economical earning, dispatching the compensation of forecast errors to the available storage systems, \textit{i.e.} \ac{bess} and \ac{hss}.

The rest of the paper is organized as follows. Section~\ref{sec:SystemArchitecture} describes the system model. Section~\ref{sec:Algorithm} introduces the optimal management algorithm. Section~\ref{sec:StudyCase} details the study case. Section~\ref{sec:Results} provides simulation results. Section~\ref{sec:Conclusions} summarizes the paper conclusions.

\section{System Model}\label{sec:SystemArchitecture}
\begin{figure}[t]
	\centering
    \includegraphics[width=1\columnwidth]{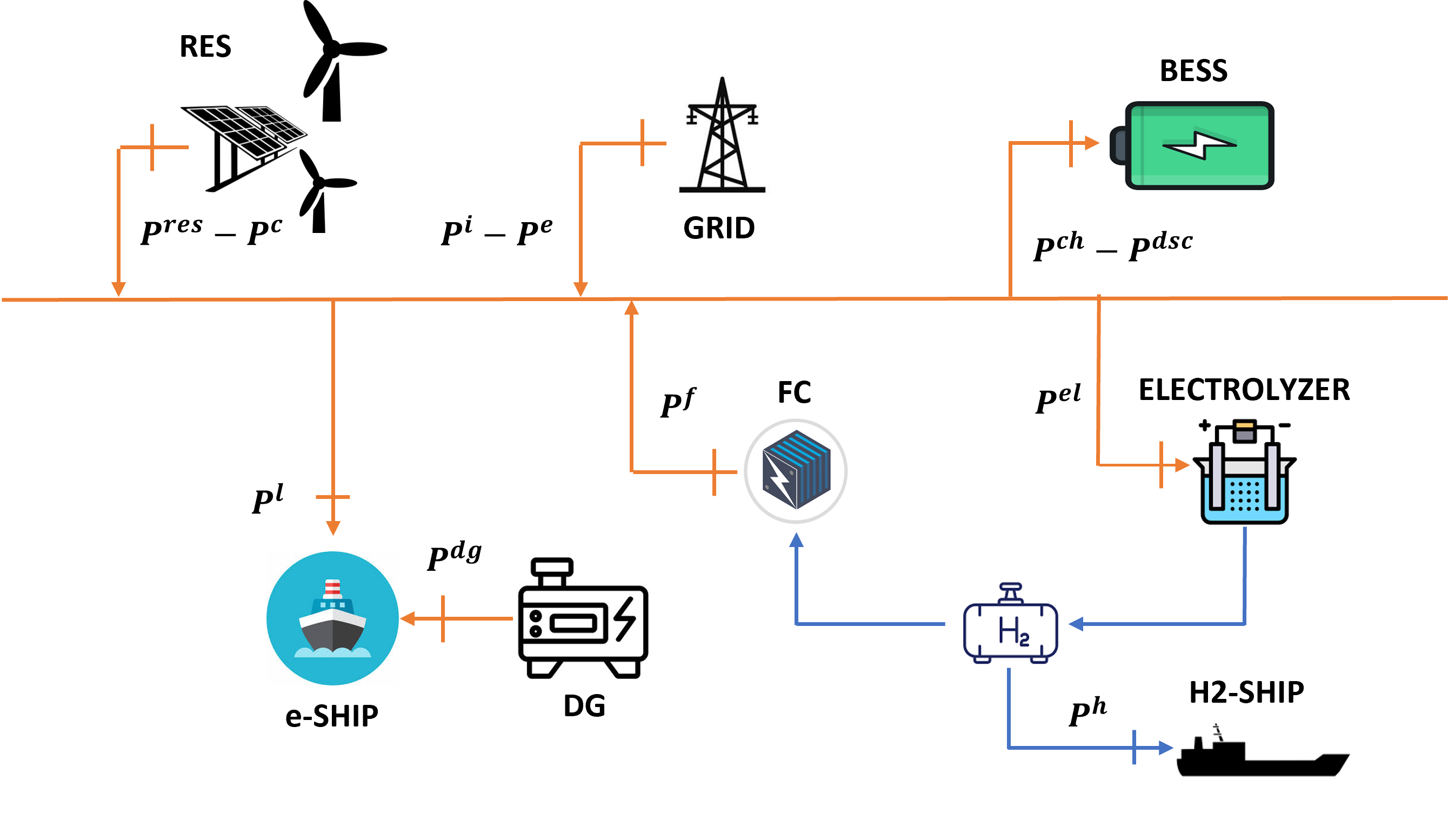}
    	\caption{System architecture.}
	\label{fig:sys_arc}
\end{figure}

The schematic architecture of the considered system is reported in Figure~\ref{fig:sys_arc}. The port is composed by two quays: one is electrified to perform \ac{ci}, one connecting \ac{ze} fueled by hydrogen. Ships connected to the electrified quay can eventually be completely or partially disconnected from the grid and satisfy their load with on-board \acp{dg}. The port is also equipped with a \ac{pv} plant, a \ac{wf}, one \ac{bess}, one \ac{hss} and a connection with the main grid. In the following, the models adopted for each system component is provided. In all models, $t$ indicates discrete-time with a sampling time of \SI{1}{\hour} and powers are considered as mean values within the sampling interval.

\subsection{Connection with the main grid}
During hour $t$, the port can import power $P^i_t$ or export power $P^e_t$. Therefore, it results that
\begin{align}
    &0 \leq P^i_t \leq \delta_t^g P_{max}^g, \label{eq: pi_lim} \\
    &0 \leq P^e_t \leq \left(1-\delta_t^g\right)P_{max}^g, \label{eq: pe_lim}
\end{align}
where $P_{max}^g$ is the rated power of the connection and $\delta^g_t$ is a binary variable.

\subsection{\ac{bess}}
The \ac{bess} is able to import or export power with the system according to its \ac{soc}. It results that
\begin{align}
    -P_{max}^b &\leq P^b_t \leq P_{max}^b, \label{eq: pb_lim} \\
    S_{t+1} &= S_{t} + \frac{1}{E^b} P^b_{t}, \label{eq:soc} \\
    S_{min} &\leq S_t \leq S_{max}, \label{eq: soc_lim}
\end{align}
where: $P^b_t$ is the power exchanged by the \ac{bess} (assuming that positive values indicate power import); $P_{max}^b$ is the \ac{bess} nominal power; $S_t$ [p.u.] is the battery \ac{soc}; $S_{min}$ and $S_{max}$ are minimum and maximum \acp{soc} to be respected for an optimal behaviour of the \ac{bess}; $E^b$ [Wh] is the battery capacity.

\subsection{\ac{hss}}
In this paper, hydrogen quantity is measured as equivalent energy, according to the transformation \SI{1}{\mega\watt\hour} = \SI{30}{\kilo\gram}. Hydrogen storage is charged by the \ac{el} and discharged by the demand of \ac{ze} and the consumption of the \ac{fc} to re-generate electrical power. The \ac{el} and the \ac{fc} have a technical operation minimum when they are switched on and can work simultaneously. Therefore, the \ac{hss} can be represented by the following equations:

\begin{align}
    &P^{el}_{min}\delta^{el}_t \leq P^{el}_t \leq P_{max}^{el}\delta^{el}_t, \label{eq: pel_lim} \\
    &P_{min}^f\delta^f_t \leq P^f_t \leq P_{max}^f\delta^f_t, \label{eq: pf_lim} \\
    &H_{t+1} = H_{t} + \frac{1}{E^h} \left( \eta^{el}P^{el}_{t} - \frac{P^f_{t}}{\eta^f} - P^h_{t} \right), \label{eq:soh} \\
    &0 \leq H_{t} \leq 1, \label{eq: soh_lim}
\end{align}
where: $P^{el}_t$ is the power consumed by the \ac{el}; $P^{f}_t$ is the power generated by the \ac{fc}; $P^h_t$ is the hydrogen required by \ac{ze}; $P_{min}^{el}$, $P_{max}^{el}$, $P_{min}^f$ and $P_{max}^{f}$ are their minimum and maximum power limits, respectively; $H_t$ [p.u.] is the \ac{soh};  $E^h$ [Wh] is \ac{hss} capacity; $\eta^{el}$ and $\eta^f$ are the efficiencies of \ac{el} and \ac{fc}, respectively; $\delta_t^{el}$ and $\delta_t^{f}$ are binary variables indicating the on/off status of \ac{el} and \ac{fc}, respectively.  

\subsection{Shore-Connection}
The power required from ships connected to the electric quay is indicated with $P^l_t$. The power eventually generated by on-board \acp{dg} is indicated with 
$P^{dg}_t$ and it must satisfy the following constraint:
\begin{equation}
0 \leq P^{dg}_t \leq P^l_t \label{eq: pdg_lim}. 
\end{equation}

\subsection{\ac{res}}
The total power generated by \ac{res} is indicated with $P_t^{res}$, whereas $P^c_t$ indicates the curtailment, which must be such that
\begin{equation}
0 \leq P^c_t \leq P^{res}_t. \label{eq: pc_lim}
\end{equation}

\subsection{Power Balance, Operational Costs and Available Data}
During every hour $t$, the following power balance needs to be matched:
\begin{equation}
    P^{res}_t + P^i_t + P^{dg}_t + P^f_t = P^l_t + P^e_t + P^b_t + P^{el}_t + P^c_t, \label{eq:powerbalance}
\end{equation}
and the port economical earning is:
\begin{equation}
J_t = c^{e}_t P^e_t + c^{ci}_t(P^{l}_t-P^{dg}_t) + c^h_t P^h_t - c^i_t P^i_t - c^c_t P^c_t
\end{equation}
where: $c^i_t$ is the energy purchase price, $c^e_t$ is the energy sell price; $c^c_t$ is the \ac{res} curtailment penalty; $c^{ci}_t$ is the tariff applied by the port to the ships in shore-connection; $c^{h}_t$ is the tariff applied by the port to \ac{ze} for hydrogen. Curtailment penalty can be applied indirectly as loss of the \ac{res} self-consumption remuneration. Concerning $c^{ci}_t$, the hypothesis is that the use of on-board \acp{dg} is discouraged by taxes that makes their generation cost $c^{dg}$ higher than the energy purchase price. As a consequence, the port has the possibility to set $c^{ci}_t<c^{dg}$, making the activation of \ac{dg} an eventual drawback, both for ships owners and for the port, which would lose an economical income equal to $c^{ci}P^{dg}_t$.

The general objective of this paper is to maximize the port economical earning, assuming that at hour $t$, given a time horizon $T$, the following data are available:
\begin{itemize}
    \item a forecast profile of \ac{res} generation $\{\hat{P}^{res}_{t+k}\}_{k=0}^{T-1}$ with an associated confidence interval $\Delta^{res}_{t+k}$, such that $|P^{res}_{t+k}-\hat{P}^{res}_{t+k}|\leq \Delta^{res}_{t+k}$;
    \item a scheduled profile of \ac{ze} hydrogen consumption $\{P^h_{t+k}\}_{k=0}^{T-1}$;
    \item a scheduled profile of the electrical load $\{P^l_{t+k}\}_{k=0}^{T-1}$;
    \item the current \ac{soc} $S_t$ and \ac{soh} $H_t$;
    \item all energy prices from time $t$ to time $t+T-1$.
\end{itemize}

\section{Optimal Management Algorithm}\label{sec:Algorithm}
At hour $t$, we will indicate with $k=0,1,\ldots,T-1$ the time sequence $t,t+1,\ldots,t+T-1$. 

The uncertainties on the \ac{res} profile forecasts are represented as follows:
\begin{equation}
    P^{res}_k = \hat{P}^{res}_k+\varepsilon_k,\quad \forall k \in [0,T-1] \label{eq:res}
\end{equation}
where the forecast error $\varepsilon_k=P^{res}_{k}-\hat{P}^{res}_{k}$ is supposed to be a normally distributed zero-mean white sequence with standard deviation $\sigma^{res}_k=\Delta^{res}_{k}/3$, so that $\mathbf{P}(|\varepsilon_k|)\leq 0.997$.

 In order to keep always satisfied \eqref{eq:powerbalance}, we assume to split out the \ac{res} forecast error between \ac{bess} and the \ac{hss}. Therefore, the powers to be exchanged by \ac{bess}, \ac{el} and \ac{fc} are set as follows:
\begin{align}
    &P^b_k=\hat{P}^b_k+\alpha^b_k\varepsilon_k \quad &\forall k \in [0,T-1], \label{eq:pb_plus_error}\\
    &P^{el}_k=\hat{P}^{el}_k+\alpha^{el}_k\varepsilon_k \quad &\forall k \in [0,T-1], \label{eq:pel_plus_error}\\
    &P^f_k=\hat{P}^f_k-\alpha^f_k\varepsilon_k \quad &\forall k \in [0,T-1], \label{eq:pf_plus_error}
\end{align}
where $\alpha^b_k$, $\alpha^{el}_k$ and $\alpha^f_k$ are \textit{partecipation factors}, which must satisfy the following conditions: 
\begin{align}
    &\alpha^b_k + \alpha^{el}_k + \alpha^f_k = 1 \quad &\forall k \in [0,T-1], \label{eq:alphas}\\
    &0\leq \alpha^b_k \leq 1 \quad &\forall k \in [0,T-1], \label{eq:ab}\\
    &0\leq \alpha^{el}_k \leq \delta^{el}_k\quad &\forall k \in [0,T-1],\\
    &0\leq \alpha^{f}_k \leq \delta^{f}_k\quad &\forall k \in [0,T-1]. \label{eq:ael_af}
\end{align}
From \eqref{eq:soc}, \eqref{eq:soh}, and \eqref{eq:pb_plus_error}--\eqref{eq:pf_plus_error}, it follows that $P^b_k$, $P^{el}_k$, $P^f_k$, $S_k$ and $H_k$ are normally distributed random variables:
\begin{align}
    &P^{\nu}_k \sim \mathcal{N}\left( \hat{P}^{\nu}_k,\psi^{\nu}_k\right), \ \nu = b, el, f  \label{eq:normal1} \\
    &S_k \sim \mathcal{N}\left( \hat{S}_k,\psi^{S}_k \right), \quad H_k \sim \mathcal{N}\left( \hat{H}_k,\psi^{H}_k \right), \label{eq:normal2}
\end{align}
where, $\forall k \in [0,T-1]$:
\begin{equation}
    \psi^{\nu}_k = \left(\alpha^{\nu}_k \sigma^{res}_k\right)^2, \qquad \nu = b, el, f,
    \label{eq:var_b_el_f}
\end{equation}
and, $\forall k \in [1,T]$:
\begin{align}
    &\hat{S}_k = {S}_0 + \frac{1}{E^b} \quad \sum^{k-1}_{j=0}\hat{P}^b_j, \label{eq:hSk}\\
    &\hat{H}_k = {H}_0 + \frac{1}{E^h} \quad \sum^{k-1}_{j=0} \left(\eta^{el}\hat{P}^{el}_j-\frac{\hat{P}^f_j}{\eta^f}-P^h_j\right), \label{eq:hHk} \\
    &\psi^S_k = \left(\frac{\eta^b}{E^b}\right)^2 \quad \sum^{k-1}_{j=0}\left(\alpha^b_j \sigma^{res}_j\right)^2, \label{eq:var_s}\\
    &\psi^H_k = \left(\frac{1}{E^h}\right)^2 \quad \sum^{k-1}_{j=0}\left(\left(\eta^{el}\alpha^{el}_j + \frac{\alpha^f_j}{\eta^f} \right) \sigma^{res}_j\right)^2. \label{eq:var_h}
\end{align}

Equations \eqref{eq:hSk}--\eqref{eq:var_h} hold true under the assumption that forecasting error at hour $k$ is independent of a forecasting error at hour $k' \neq k$, and considering that the initial \ac{soc} ${S}_0$ and \ac{soh} $H_0$  are available.

Random variables in \eqref{eq:normal1}--\eqref{eq:normal2} must satisfy constraints \eqref{eq: pb_lim},\eqref{eq: soc_lim}--\eqref{eq: pf_lim} and \eqref{eq: soh_lim}. To deal with their randomness, chance constraints are introduced. For a generic normally distributed random variable $x \sim \mathcal{N}\left( \hat{x},\sigma^2 \right)$ the chance constraint:
\begin{equation}
    \mathbf{P}\left(\underline x \leq x \leq \overline x \right) \geq 1-\beta
\end{equation}
can be written with the following inequalities:
\begin{equation}
    \hat{x} + \theta \sigma \leq \overline x, \quad
    -\hat{x} + \theta \sigma \leq \underline x
\end{equation}
where $\theta=\sqrt{2}$erf$^{-1}\left(1-2 \beta \right)$ \cite{Conte:2020}. Therefore, we obtain that, $\forall k \in [0,T-1]$:
\begin{align}
    \hat{P}^{\nu}_k + \theta \alpha^{\nu}_k \sigma^{res}_k \leq P^{\nu}_{max}, &\quad -\hat{P}^{\nu}_k + \theta \alpha^{\nu}_k \sigma^{res}_k \leq P^{\nu}_{min}, \label{eq:chance_p}\\
    \hat{S}_k + \theta \sqrt{\psi^S_k} \leq S_{max}, &\quad -\hat{S}_k + \theta \sqrt{\psi^S_k} \leq S_{min}, \label{eq:chance_s} \\
    \hat{H}_k + \theta \sqrt{\psi^H_k} \leq 1, &\quad -\hat{H}_k + \theta \sqrt{\psi^H_k} \leq 0, \label{eq:chance_h}
\end{align}
where $\nu=b,el,f$.

\begin{figure}[t]
	\centering
    \includegraphics[width=1\columnwidth]{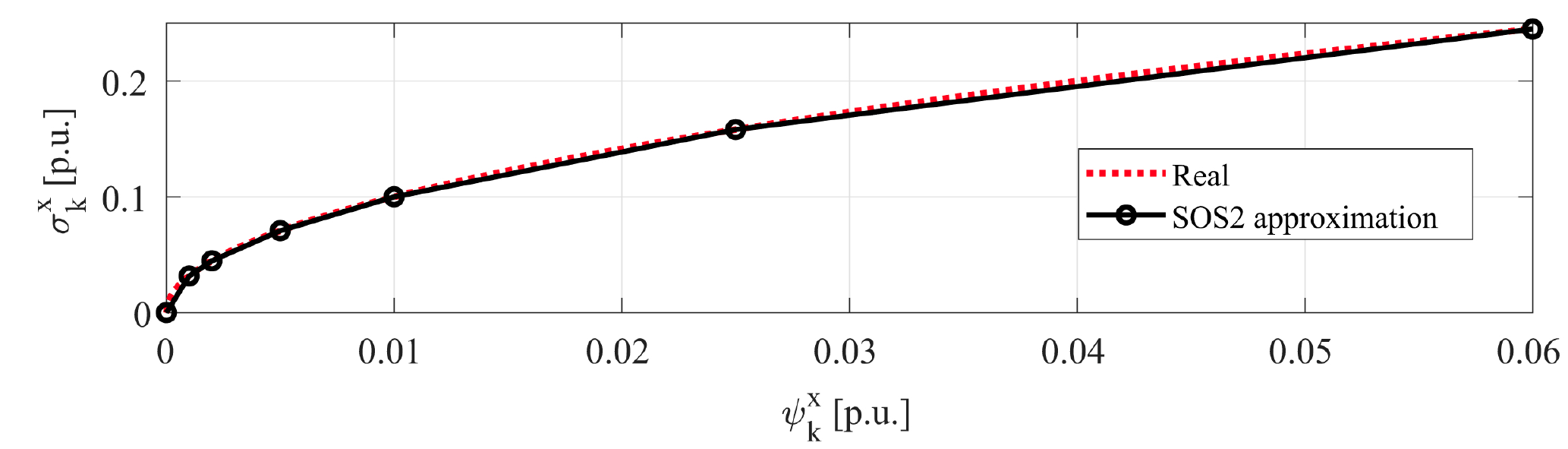}
    	\caption{Piecewise linearization of the square root function applied to constraints \eqref{eq:chance_s} and \eqref{eq:chance_h}}
	\label{fig:sos}
\end{figure}

Constraints \eqref{eq:chance_s}--\eqref{eq:chance_h} are nonlinear and non-convex because of the presence of the square root function. To cope with this problem, a piecewise linearization is applied as reported in Figure~\ref{fig:sos}. Such an approximation can be included in the optimization problem by introducing a set of binary variables or the so called \ac{sos2} variables \cite{Conte:2019}. Details are not provided in this paper for space lacking.

%
%


The proposed optimal management algorithm is based in the \ac{mpc} \textit{receding horizon principle}, that consists in solving an optimization problem defined for a given time horizon at time $t$, applying the first element of the solution trajectories and repeat the same at time $t+1$.
In our case, the optimization problem is formulated as follows:
\begin{equation}
\begin{aligned}
    &\max_{\{X_k\}} \sum^{T-1}_{k=0} J_k \\
    &X_k=\left[P^i_k, P^e_k, P^{dg}_k, P^c_k, \hat{P}^b_k, \hat{P}^{el}_k, \hat{P}^f_k, \alpha^b_k, \alpha^{el}_k, \alpha^f_k \right]^\top \label{eq:cost_function}
\end{aligned}
\end{equation}
subject to:
\eqref{eq: pi_lim}--\eqref{eq: pe_lim}, \eqref{eq: pdg_lim}--\eqref{eq:powerbalance}, \eqref{eq:alphas}--\eqref{eq:ael_af}, \eqref{eq:hSk}--\eqref{eq:var_h}, \eqref{eq:chance_p}--\eqref{eq:chance_h}.

The  variables reported in vector $X_k$ includes only the control variables to be applied to the system. Actually, also $\delta_k^g$, $\delta_k^el$ and $\delta_k^f$ and \ac{sos2} variables belong to  $X_k$. Since \eqref{eq:var_s}--\eqref{eq:var_h} are quadratic with respect to $\alpha^b_k$, $\alpha^{el}_k$, and $\alpha^f_k$, \eqref{eq:cost_function} results to be a \ac{miqcp}.




\section{Study Case}\label{sec:StudyCase}
The considered study case is an hypothetical Smart Port located in the area of the Genova harbour in Italy. Table~\ref{tab:smartport_parms} reports the values assigned to model parameters. All components was sized using the software Hybrid Optimization of Multiple Energy Resources (HOMER) \cite{Homer}, which has the capability to simulate the various combination of technologies in a microgrid taking into account the shared energy and the costs of the system. The design was carried out by selecting among a set of sizes for each component, established by the user, and minimizing the \ac{lcoe}.

The yearly electrical demand of a the shore-connection was defined based on \cite{GM2021}, which provides an estimation of the total electrical load of roll-on roll-of passengers (Ro-Pax) ships moored at a quay in the Genova port, in 2019. The same was done to define yearly demand of hydrogen from \ac{ze}, using the aforementioned equivalence between hydrogen and electrical energy and assuming that they are equipped with on-board \acp{fc} with an efficiency of 60\%. We assumed that only one Ro-Pax ship and one \ac{ze} Tanker at a time can be moored at the corresponding quay. 

To size Wind and \ac{pv} power plants, wind speed and solar irradiation in the area of Genova was taken from the NASA Prediction of Worldwide Energy Resource database \cite{nasaRES}. To establish the maximal size of the \ac{pv} power plant, we assumed the availability of an area of \SI{123800}{\meter^2}, estimated by Authority of the port system in \cite{deasp}, which, with an expected efficiency of \SI{0.1}{\kilo\watt/\meter^2}, returned a potential peak power of \SI{12.38}{\mega\watt}. HOMER optimization returned a nominal power for the \ac{pv} power plant of \SI{4}{\mega\watt} and a nominal power for the \ac{wf} of \SI{11.34}{\mega\watt}, composed by 14 \SI{810}{\kilo\watt} Wind Turbines.
Efficiencies and potential sizes of \ac{bess}, \ac{fc}, \ac{el}, and \ac{hss} was taken from datasheets of available commercial devices.


\begin{table}[t]
		\centering
		\caption{Study case port parameters.}
		\label{tab:smartport_parms}
		{\small
		\begin{tabular}{lcc}
		\toprule
		Parameter & Symbol  & Value\\
		\midrule
        Main Grid Connection Nominal Power & $P^g_{max}$ & \SI{16}{\mega\watt} \\
        \ac{bess} Nominal Power & $P^b_{max}$ & \SI{1}{\mega\watt} \\
        \ac{bess} Capacity & $E^b$ & \SI{2.5}{\mega\watt\hour} \\
        \ac{bess} Round Trip Efficiency & $\eta^b$ & \SI{95}{\percent}  \\
        \ac{hss} Capacity & $E^h$ & \SI{2000}{\kilo\gram} \\
        \ac{hss} Nominal Load & $-$ & \SI{19.95}{\kilo\gram/\hour} \\
        \ac{fc} Nominal Power & $P^{f}_{max}$ & \SI{0.25}{\mega\watt} \\
        \ac{fc} Technical Minimum & $P^{f}_{min}$ & \SI{0.025}{\mega\watt} \\
        \ac{fc} Efficiency & $\eta^{f}$ & \SI{60}{\percent} \\
        \ac{el} Nominal Power & $P^{el}_{max}$ & \SI{1.5}{\mega\watt} \\
        \ac{el} Technical Minimum & $P^{el}_{min}$ & \SI{0.15}{\mega\watt} \\
        \ac{el} Efficiency & $\eta^{el}$ & \SI{70}{\percent} \\
        Shore-Connection Nominal Power & $-$ & \SI{11.96}{\mega\watt} \\
        \ac{pv} Plant Nominal Power & $-$ & \SI{4}{\mega\watt} \\
        \ac{wf} Nominal Power & $-$ & \SI{11.34}{\mega\watt} \\
		\bottomrule
		\end{tabular}
        }
\end{table}

\section{Simulation Results}\label{sec:Results}
To test the performance of the proposed control algorithm, one week has been simulated, using the data from the midnight of August 3rd to the midnight of August 11th, 2019. The adopted energy prices are reported in Table~\ref{tab:prices}. Notice that the price of hydrogen ($c^h_t$), which is hard to estimate for the future, has been set equal to the one of cold ironing $c^{ci}_t$. The following control parameters have been set: $T=\SI{12}{\hour}$, $\beta=0.05$, $S_{max}=\SI{0.9}{p.u.}$ and $S_{min}=\SI{0.1}{p.u.}$.

The \ac{res} forecast profiles has been taken from \cite{nasaRES}, whereas real profile has been computed using \eqref{eq:res}, with a resulting standard deviation $\sigma^{res}=\SI{0.273}{\mega\watt}$. The control algorithm has been implemented in MATLAB, integrated with \ac{gams} to write the optimization problem, solved by the SCIP solver.  

\begin{table}[t]
		\centering
    \includegraphics[width=1\columnwidth]{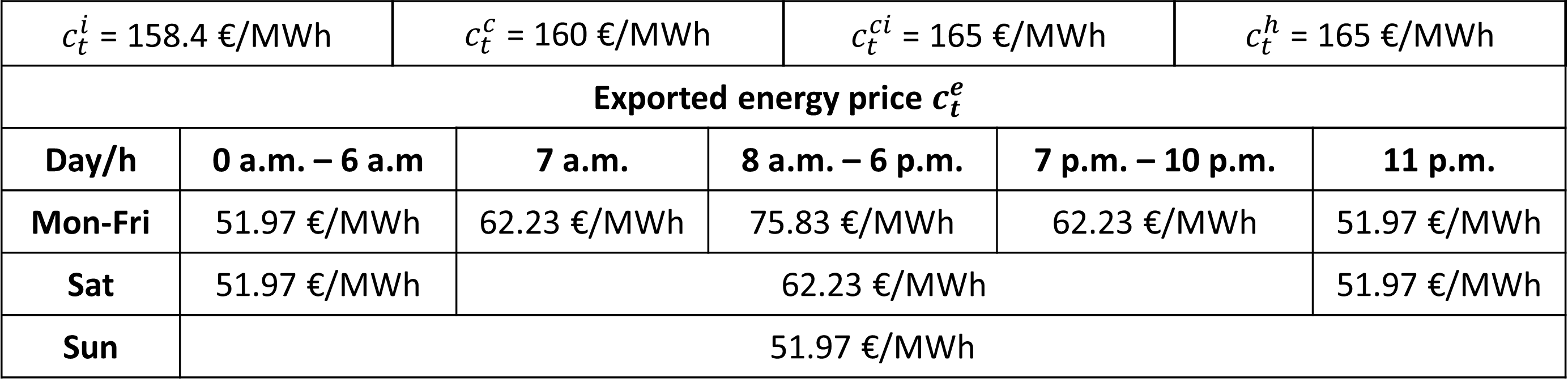}
    	\caption{Simulation scenario: energy prices.}
	\label{tab:prices}
\end{table}



\begin{figure}[t]
	\centering
    \includegraphics[width=1\columnwidth]{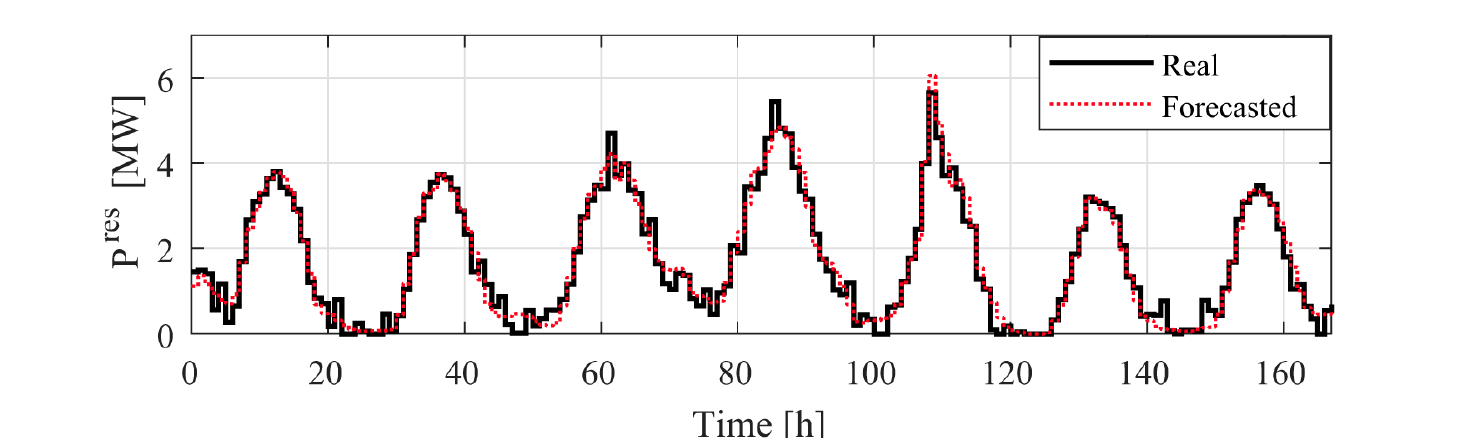}
    	\caption{\ac{res} profiles}
	\label{fig:res}
\end{figure}

\begin{figure}[t]
	\centering
    \includegraphics[width=1\columnwidth]{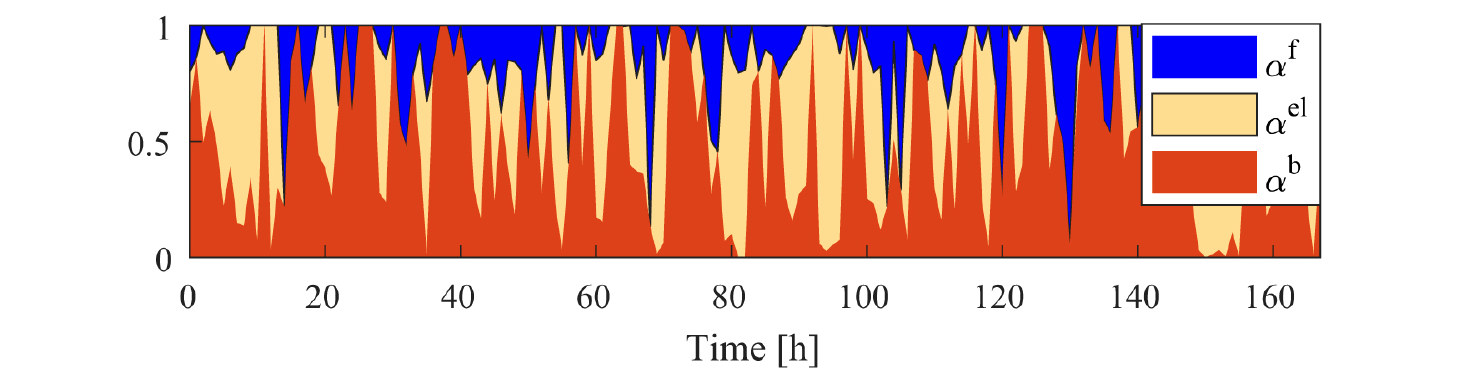}
    	\caption{\ac{res} compensation participation factors profiles.}
	\label{fig:alpha}
\end{figure}

\begin{figure}[t]
	\centering
    \includegraphics[width=1\columnwidth]{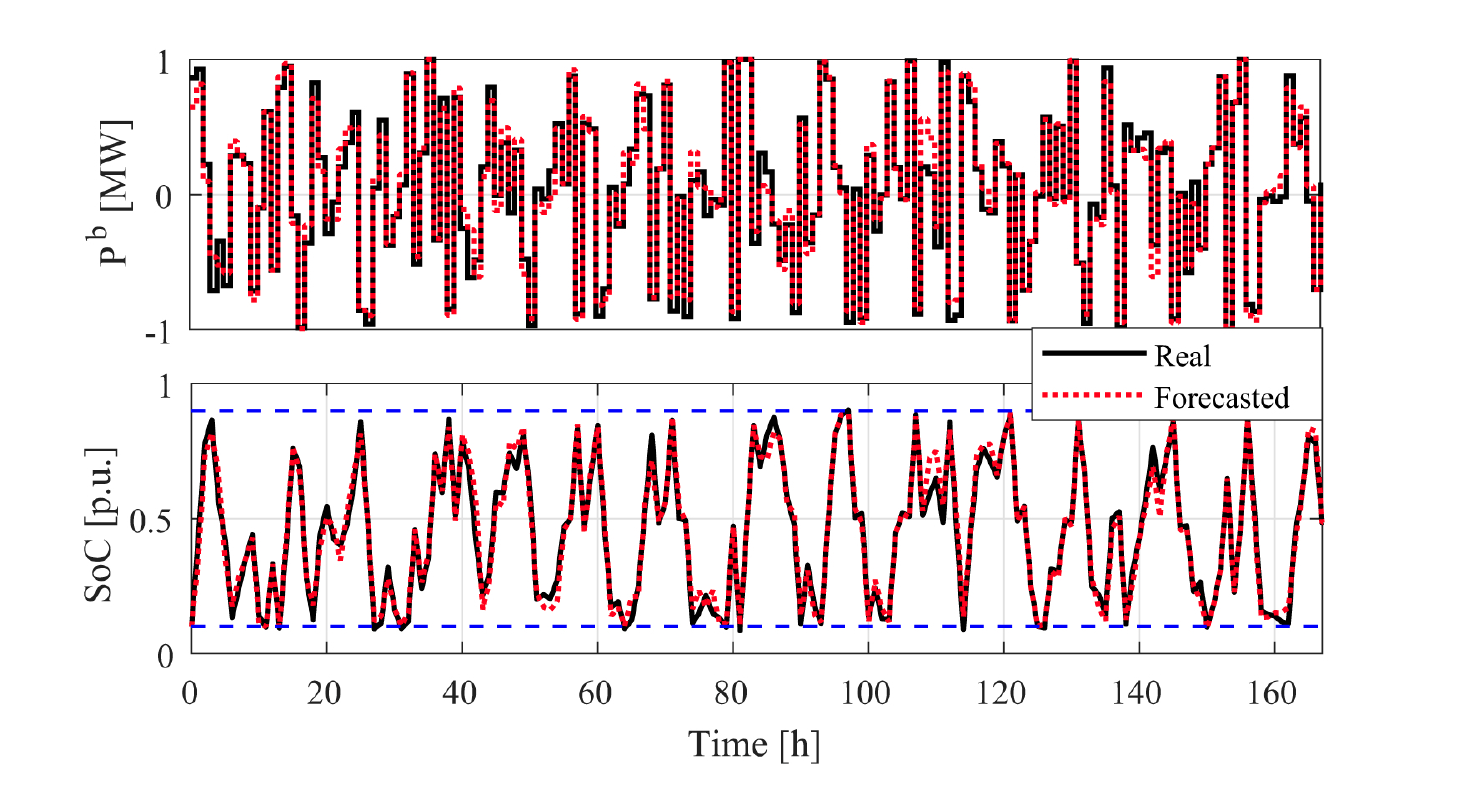}
    	\caption{\ac{bess} exchanged power (top) and \ac{soc} (bottom).}
	\label{fig:bess}
\end{figure}

\begin{figure}[t]
	\centering
    \includegraphics[width=1\columnwidth]{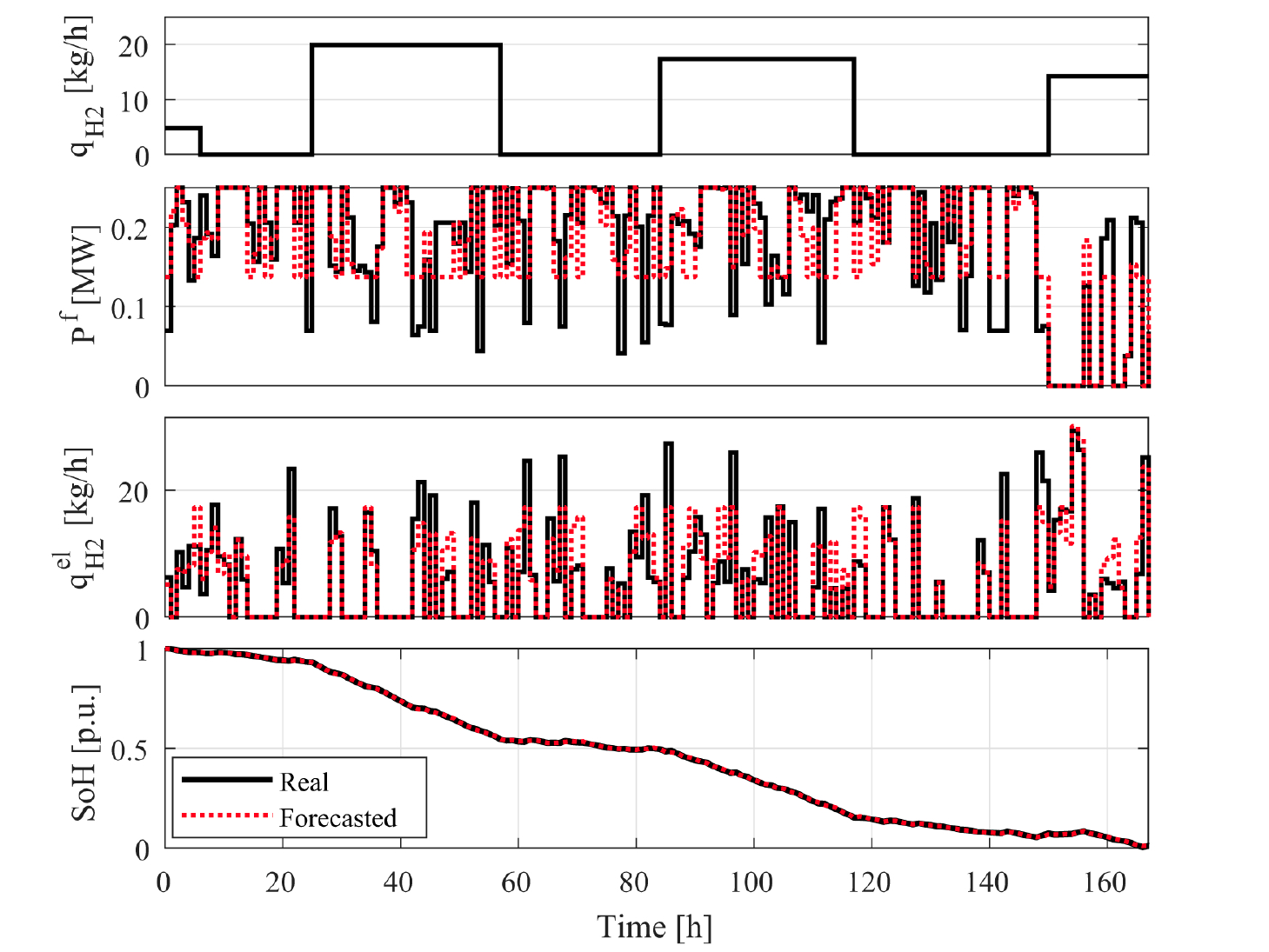}
    	\caption{\ac{ze} Hydrogen load (top), power profiles of \ac{fc} (mid-top) and \ac{el} (mid-bottom), and \ac{soh} profile (bottom).}
	\label{fig:h2}
\end{figure}

\begin{figure}[t]
	\centering
    \includegraphics[width=0.95\columnwidth]{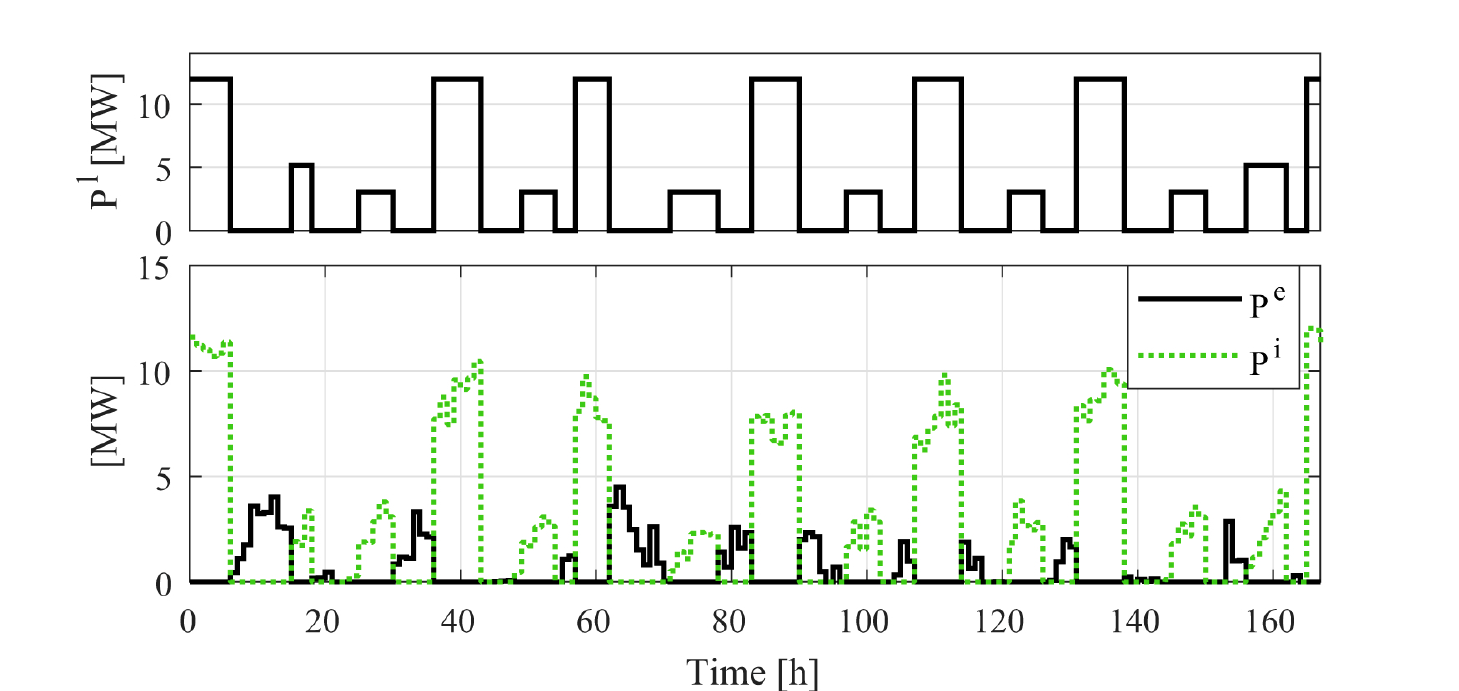}
    	\caption{Electric ships load (top) and port import and export power profiles (bottom).}
	\label{fig:l-i-e}
\end{figure}

Figures~\ref{fig:res}--\ref{fig:l-i-e} show the obtained simulation results. Notice first that $P^c_t$ and $P^{dg}_t$ are not shown since they resulted to be always zero, meaning that no curtailment has been performed and the use of on-board \acp{dg} has been avoided. None of the constraints has been violated, meaning that both the electrical and the hydrogen loads have been fully satisfied. In Figure~\ref{fig:alpha} we can observe how the algorithm has decided the contribution to the compensation of the \ac{res} forecast errors among \ac{bess}, \ac{fc} and \ac{el} and their power profiles in Figure~\ref{fig:bess} and Figure~\ref{fig:h2} we can observe the effect of these contributions, with a deviation of the real profile from the forecasted one.

We finally remark that the total earning obtained with the week operations has resulted to be equal to \SI{45641}{\eur}.

\section{Conclusions}\label{sec:Conclusions}
A management strategy for economically optimizing the operational costs of a green multi-energy smart port performing cold-ironing and hydrogen supplying has been proposed. The control approach, based on \ac{mpc}, takes into account the uncertainty of \ac{res} generation exploiting the two storage systems installed in the port: a battery and a hydrogen tank coupled with an electrolyzer and a fuel cell.
Future works will consider different study cases to prove the robustness of the approach and its development, with an interaction of the proposed algorithm with a week- or month- ahead planning required especially for the hydrogen demand. 


\end{document}